\documentstyle[twoside,fleqn,espcrc2]{article}

\newcommand{\AmS}{{\protect\the\textfont2
  A\kern-.1667em\lower.5ex\hbox{M}\kern-.125emS}}

\title{
\vspace{-5.0cm}
\begin{flushright}{\normalsize RU-98-30}\\
\end{flushright}
\vspace*{2.5cm}
Lattice Chirality
\thanks{Talk at Lattice '98, Boulder Colorado, July 13-18, 1998}}

\author{Herbert Neuberger\address{Department of Physics and Astronomy, \\ 
        Rutgers University, Piscataway, NJ 08855-0849}%
        \thanks{Research supported in part by the DOE, 
grant \#DE-FG05-96ER40559.}
        }
       
\begin{document}

\begin{abstract}
The external fermion propagator and the 
internal fermion propagator in the overlap
are given by different matrices. A generic
problem (formulated by Pelissetto)
faced by all chiral, non-local, propagators
of Rebbi type is avoided in this manner. 
Nussinov-Weingarten-Witten mass inequalities
are exactly preserved. It is sketched how
to obtain simple lattice 
chiral Yukawa models
and simple expressions for covariant currents.
Going beyond my oral presentation, 
I have added to the write-up several comments
on Niedermayer's talk. His transparencies are 
available on the internet.
\end{abstract}

\maketitle

Chiral symmetry was built into the overlap from
day one. Each Weyl fermion is represented by 
an infinite tower and coherent unitary
rotations of left-towers into linear
combinations of left towers (the 
same holds for right towers) are exact 
global symmetries in vector-like gauge theories.
These are symmetries of the fermion quantum
system underlying the overlap and are realized
canonically. There are also global
$U(1)$ symmetries associated with each tower. Fermion
correlation functions are produced by taking
matrix elements of strings of creation/annihilation operators 
between a reference state and
a fermionic ground state. The ground state
depends parametrically on the gauge background,
but the reference state can be chosen not to. 
The ground state can transform 
nontrivially under
one linear combination of the $U(1)$s, 
axial-$U(1)$, but is a singlet under all other
global chiral symmetries as long as the gauge
background is smooth enough. 
This provides 
a lattice realization of 't Hooft's solution
to the $U(1)$-problem. The backgrounds where
axial-$U(1)$ is violated contain net topological
charge. In addition, a certain choice for
the external fermion propagator produces
an exact lattice realization of the 
Nussinov-Weingarten-Witten mass inequalities. 
All of the above is fully explained in
section 9 of \cite{npblong}. In other sections
evidence is provided for correct realization
of chiral anomalies. Thus, spontaneous 
chiral symmetry breakdown is not only plausible
beyond reasonable doubt, but potentially
rigorously provable.

The simplified formula obtained for the overlap
in \cite{plba} is an expression of the matrix element
of unity, the chiral determinant, $\det (D)$
with $D=(1+V)/2$. Here, $V=\gamma_5\epsilon(H)$,
with $H=\gamma_5 D_W$, where $D_W$ is the 
Wilson-Dirac operator with hopping larger than
critical. Let the size of $D_W$ be 
${\cal N}\times{\cal N}$.
$D$ is obtained by starting
from a $2{\cal N}\times 2{\cal N}$ hamiltonian
$H^\prime_2 =\pmatrix {H& 0\cr 0&-H\cr}$. By
conjugation with ${1\over \sqrt{2}}\pmatrix
{1&-1\cr 1&1}$ one gets $H_2 =
\pmatrix {0& H\cr H&0\cr}$. To get the external
fermion propagator one needs the projectors
on the positive and negative eigenspaces 
of $H_2$ (see section 5 of \cite{npblong}). In
backgrounds carrying zero lattice topological
charge they are $2{\cal N}\times {\cal N}$ 
rectangular matrices, $P_\pm (H)={1\over\sqrt{2}}
\pmatrix {\epsilon(H)& \pm 1\cr}$. The reference
state is obtained by substituting $\gamma_5$
for $H$ (One can easily work with a gauge field
dependent reference state, substituting 
$H_0$ for $H$, where $H_0$ differs from $H$
in that its hopping
parameter is below the critical
value.) Since $\epsilon (\gamma_5 )=\gamma_5$
we have $P_\pm (\gamma_5 )={1\over\sqrt{2}}
\pmatrix {\gamma_5 & \pm 1\cr}$.

Equations (5.7) and (5.19) in \cite{npblong} 
provide an expression for the propagator
$\tilde G_2$:
\begin{eqnarray}
\tilde G_2 =& P_+^\dagger (H) 
{1\over{P_+ (\gamma_5 ) P_+^\dagger (H) }} 
P_+ (\gamma_5 )= \nonumber \\
&{1\over{1+V}}
\pmatrix{1&\gamma_5\cr\gamma_5 & 1\cr}
\end{eqnarray}
Rotating back to the original frame, we get
\begin{equation}
\tilde G_2^\prime = {1\over{1+V}}
\pmatrix{1+\gamma_5 & 0\cr 0&1-\gamma_5\cr}\end{equation}
As noted in \cite{npblong} we see that only half
the modes propagate, so we have effectively
an ${\cal N}\times{\cal N}$ restricted
propagator. It
is chiral, so has only left-left and right-right
entries in the chiral basis:
$\tilde g_{RR}=\tilde g_{LL}={2\over{1+V}}$. 
(The chiral projectors ${{1\pm\gamma_5}\over 2}$
are assumed implied by the subscripts.)
However, it
violates the continuum relation 
$g_{RR}=-g_{LL}^\dagger$. To restore this
essential relation for NWW mass inequalities the ordering ambiguity
inherent in the overlap construction of fermionic
correlation functions was exploited in \cite{npblong} to obtain
another expression, given in equation (5.22):
\begin{equation}
g_{RR} =-g_{LL}^\dagger = {2\over{1+V}}-1=
{{1-V}\over{1+V}}\end{equation}
These can be assembled into a chiral, non-local
propagator of Rebbi type: 
\begin{equation}
g={{1-V}\over{1+V}}\end{equation}
In equation (2.22) of \cite{almostmassless} it was
shown that a physical expression for the
chiral condensate in a fixed gauge background
in the presence of $N_f$ flavors is 
proportional to
\begin{equation}
\det{}^{N_f} \left ( {{1+V}\over 2}  \right ) {\rm Tr} {{1-V}\over
{1+V}},\end{equation}
succinctly exhibiting the dichotomy between
internal and external propagators. The expression
would vanish in finite volumes as long as 
$N_f >1$, exactly as expected. 

Had $g$ been the
internal free fermion propagator, the induced 
effective action would have had unwanted contributions
from ghosts which would survive in the continuum
limit, as shown by Pelissetto
\cite{peli}. But, here the effective action is
given by $\det (D)$, not $\det ( g^{-1} )$, and
the free propagator $D^{-1}$ has no ghosts.

In any explicit realization of strictly massless QCD
adding a mass terms (see \cite{npblong,almostmassless}) 
is easy and the dependence on $\theta_{\rm QCD}$ can
be made explicit. Using the overlap, and making $\theta$ space
dependent, one immediately obtains a 
simple lattice chiral Yukawa model.

In \cite{geom} the definition of covariant currents
was discussed in the general chiral context
including the vector-like case. 
If that definition is applied directly
to $H$ one obtains rather ugly expressions
because of nontrivial energy denominators.
However, $H$ can always be replaced by an odd
monotonic function $f(H)$. In \cite{last} 
a method to compute $D$ was presented. It is based
on a function $F_n (z)$,
\begin{equation}
F_n (z) = {{(1+z)^n - (1-z)^n}\over
{(1+z)^n + (1-z)^n}}\end{equation}
$F_\infty (H) = \epsilon (H)$. A crucial
relation is $F_n(F_m (z))=F_{nm} (z)$. 
We learn that we can substitute $F_m (H)$
for $H$. For very large $m$ the 
energy denominators simplify and 
nice expressions for the gauge invariant chiral 
currents can be obtained. 

Using the identity
\begin{eqnarray}
&F_{2n} (z)=\nonumber\\
&{z\over n} \sum_{s=1}^n {1\over{z^2 \cos^2 {\pi\over{2n}}
(s-{1\over 2}) +\sin^2 {\pi\over{2n}} (s-{1\over 2})}}
\end{eqnarray}
in conjunction with a shifted CG solver the action of $\epsilon (H)$
on a vector can be easily evaluated \cite{last}. 
A refinement of this method has appeared in \cite{refine}. 

See \cite{luscher} for other work related to the above. 

Niedermayer, in his plenary talk, applied revisionism to the history
of exact chiral symmetry on the lattice. 
His 46 transparencies are available at 
http://pizero.Colorado.EDU/Lattice98/Planary\linebreak/Niedermayer/ 
(note unusual spelling). 
Below are comments I could fit in the remaining allotted space. 
Some of the comments apply also to the 
plenary talk given by Sharpe at ICHEP'98.

The renormalization group works by making the ultraviolet complications
of field theory emerge in the limit of infinite iteration of a
regular step. Regularity means real analyticity in momentum space (locality),
real analyticity in field space and real analyticity in the couplings.
The iteration is to be applied to a 
starting point that obeys desired symmetries and the above three-fold
analyticity. A lattice-Dirac operator that is analytic in the group
valued gauge variables cannot reproduce the disconnected nature
of the collection of all smooth continuum gauge fields over a compact
manifold. On the lattice any gauge configuration can be smoothly
deformed into any other configuration and robust exact zero modes
can appear only by going through points where the dependence on the
gauge fields is singular. Thus, a lattice model with exact chiral
symmetry and an action bilinear in the fermi fields must
violate the condition of analyticity in field space. Therefore, 
an ideal lattice fixed point action
for massless $QCD$ should not be bilinear in the fermi fields. 
The understanding of the limitations of QCD 
``perfect actions'' (which are not ideal fixed points) 
is incomplete: to how many
loops does ``perfection'' hold, what is the explicit form
of the perfect fermionic action, 
do any solutions to ``perfect fixed point'' equations exist,
how unique is a solution ? 
Without analyticity in field space
there is a danger of loosing universality: for example,
we can now obtain arbitrary critical
exponents in mean-field. To approach
the desired continuum limit the singularities
in field space have to decouple from long distance physics. 
In an e-mail message to L{\" u}scher (Jan. 23, 1998)
I suggested that in the vicinity of certain singularities
in field space also the locality of the overlap Dirac operator
might be lost, but that this would be avoided for actions
with a plaquette value restrictively 
bounded from below (see the 30-th transparency). 

The overlap-Dirac operator is {\it equivalent} to the overlap presented
in \cite{npblong}. Thus, global chiral symmetries are guaranteed,
as they would in any theory in which there is no direct coupling between
left and right Weyl fermions. These symmetries are masked by the effects
of integrating out an infinite number of fermi and pseudo-fermi fields. 
The simpler $1+V$ formula appeared first in \cite{plba}, one year ago, 
and could
not have been prompted by work of Hasenfratz et al. The topological
properties were known from \cite{npblong} and were re-derived in the
newer form of \cite{plba}. A mass term was introduced in \cite{npblong}
and explicit expressions were written down in \cite{almostmassless}.
Replacing $D$ by $1+V$, one can obtain the
equations on the 16-th transparency from \cite{almostmassless}. 

For chiral gauge theories \cite{npblong} one needs the projectors on the
positive and negative eigenspaces of $H$ itself, not only $H_2$. 
So long as the gauge fields have zero topology in the sense of the
overlap the projectors will be ${\cal N}\times{{\cal N}\over 2}$
matrices 
representing the action of ${{1\pm\epsilon(H)}\over 2}$.
However, unlike for $H_2$, the second factor, ${{\cal N}\over 2}$,
will change when the ``instanton number'' is non-zero.
Explicit formulae for these projectors require the eigenvectors
of $H$ and the sign of the related eigenvalues \cite{npblong}. This
is equivalent to spectrally resolving $\epsilon (H)$. On the 19-th
transparency $\epsilon(H)\equiv \gamma_5 (1+V-1)$ 
was denoted by $\hat \gamma_5$ a notation
credited to Hasenfratz, Niedermayer and L{\" u}scher. 
The eigenvectors and the associated propagator \cite{npblong} 
appear on the 44-th transparency. The formulae
for topological charge are those of the overlap. As explained in 
\cite{npblong} and its predecessor, NPB412(1994)574, the overall
phase ambiguity is directly related to anomalies. A fundamental connection
between these phases and the algebraic conditions for continuum
anomaly cancelation was described in \cite{geom}. The importance
of the $\pm 1$ modes of $V$ emphasized at the bottom of the 21-st
transparency was first stressed in comment (c) of hep-lat/9805027.
The introduction of a $\theta$-parameter (topic of the 22-nd transparency)
was first discussed in section 9 of \cite{npblong}. The absence of order
$a$ effects is trivial in view of chiral symmetry and was explicitly
mentioned in a footnote on page 110 of PLB399(1997). 
The local properties of the free system were checked in PLB302(1993)62.
That the free system is related to GW was mentioned 
in \cite{almostmassless} after eq. (2.24) there. \cite{almostmassless} 
was posted
on hep-lat in October 1997. Using the Foerster,
Nielsen, Ninomiya mechanism (46-th transparency)
was emphasized in \cite{npblong} and tested
in two dimensions thereafter.

\end{document}